# The problem of what exists*

P.C.W. Davies

Australian Centre for Astrobiology, Macquarie University, New South Wales, Australia 2109

**Abstract**

Popular multiverse models such as the one based on the string theory landscape require an underlying set of unexplained laws containing many specific features and highly restrictive prerequisites. I explore the consequences of relaxing some of these prerequisites with a view to discovering whether any of them might be justified anthropically. Examples considered include integer space dimensionality, the immutable, Platonic nature of the laws of physics and the no-go theorem for strong emergence. The problem of why some physical laws exist, but others which are seemingly possible do not, takes on a new complexion following this analysis, although it remains an unsolved problem in the absence of an additional criterion.

## 1. Background

The puzzle of why the universe consists of the things it does is one of the oldest problems of philosophy. Given the seemingly limitless possibilities available, why is it the case that atoms, stars, clouds, crystals, etc. are "chosen" to exist in profusion in preference to, say, pulsating green jelly or pentagonal chain mail? A related question is why the entities that do exist conform to the particular physical laws that they do as opposed to any other set of laws one might care to imagine. Physicists have mostly ignored this problem, content to accept the observed physical systems and their specific laws as "given," and preferring to concentrate on the job of elucidating them. Notable exceptions were Einstein, who famously remarked that he wanted to know whether "God had any choice" in the nature of his creation, and Wheeler, whose rhetorical question "How come existence?" provided the basis for a series of speculative papers (Wheeler, 1979, 1983, 1988, 1989, 1993).

Recently, however, theoretical physicists and cosmologists have been giving increasing attention to the problem of "what exists". In part this stems from the growing interest in unification, especially string/M theory, and the concomitant sharp disagreements about uniqueness (see, for example, Danielsson, 2001). Meanwhile, the popularity of multiverse cosmological models has prompted a dramatic reappraisal of the very concept physical existence.

The issues are clarified in Fig. 1. The picture shows three sets separated by two boundaries, *A* and *B*. The middle region is the set of *all things that observers can in principle observe*. (At the moment, of course, humans have actually observed only a fraction of what is "out there".) The set delineated by *A* can be a subset of *all that exists*. Then there is a bigger set, containing the other two: the set of all that *can* exist. The principal question I shall address in this paper is how one might determine the location of the boundaries *A* and *B*.

A common claim among string/M theorists is that *A* coincides with *B*; that is, the set of all that can be observed is the same as the set of all that exists (Danielsson, 2001). (By this, I refer to all *fundamental entities* and laws that can in principle be observed. I am excluding unique or unusual macroscopic objects that might be unobservable because they form beyond an event or particle horizon, for example.) This claim derives from the hope that the final unified theory will be in some sense unique, so that it describes a single possible world consistent with the known facts. A very different claim is made, however, by another group of theoretical physicists and cosmologists, who postulate a multiverse of different universes, and invoke an observer selection effect to define boundary *A*. For example, in the string theory landscape model of Susskind (2005), there are $> 10^{500}$ possible vacuum states of the theory, each with distinct low-energy physics. When the landscape concept is combined with eternal inflation (see for example Linde, 1990), there is a mechanism to populate the landscape with actual universes. The vast majority of these universes would be inconsistent with the emergence of life and observers, and so would go unseen. In that case, "what is observed" is a minute subset of all that exists, and boundary *A* is defined by the criteria necessary for the emergence of life and observers (Barrow and Tipler, 1986). We don't have a very clear idea of what those criteria are (the existence of carbon, water and stable stars are often cited); however, this is a purely scientific matter that could one day be settled.

It is not my intention to review the arguments for and against the competing positions in regard to boundary *A*, since they have been thoroughly discussed elsewhere (see, for example, Susskind 2005). Rather I wish to focus the discussion on boundary *B*, and specifically on three attempts to explain it (or explain it away). Boundary *B*, remember, separates that which exists from that which is possible but in fact non-existent.

1. *Unique universe*

The claim is sometimes made that boundary *B* does not exist, that is, the set of all that exists is the set of all that *can* exist. Expressed differently, this claim says that the universe must exist necessarily as it is or, to paraphrase Einstein, that "God had no choice" in its nature because there is only one possible world. The justification for this viewpoint rests on a belief that a truly unified theory of physics would have no free parameters, and its mathematical form would be so tightly constrained by logical self-consistently that it would be unique, and possess a unique "solution" representing the physical state of the universe. (There may of course be flexibility within this unique state on account of quantum indeterminism.)

It is easy to demonstrate that the foregoing claim is false in the form stated. A traditional practice among theoretical physicists is to construct self-consistent mathematical models that are simpler than the real universe. For example, the Thirring model in quantum field theory, mini-superspace models of quantum gravity, Boltzmann gases. These models offer impoverished descriptions of reality, and are studied because they capture in some useful way a restricted feature of nature. Although they are not serious contenders for descriptions of the real world, nevertheless they describe *possible worlds*. And it is hard to see any limit on the number of different artificial mathematical models of this sort. So the universe quite clearly could have been otherwise in a seemingly limitless number of ways.

It might be possible to establish a weaker claim: that there is a unique universe consistent with all current observations. The additional criterion seems to be implicit in discussions of uniqueness in final theories, although the restriction is often quietly dropped when applying it to the problem of boundary *B*. Other possible additional criteria come to mind. It could be that the observed universe is the uniquely *simplest* universe that is nevertheless rich enough to permit the existence of life. In that case boundaries *A* and *B* coincide, but it remains the case that the common boundary delineates only a small subset of all that *can* exist.

2. *The best of all possible worlds*

Leibniz (1697) was among the first to recognize that the world could have been otherwise, that the arrangement of matter in time and space could have varied in an endless number of ways. He even considered the possibility of a multiverse containing other regions of space and time, although he subsumed them all under his definition of "World." Famously, Leibniz claimed that ours is *the best of all possible Worlds*. So Leibniz recognized the existence of boundary *B* (as far as I know he made no attempt to discuss boundary *A*), and he defined it in terms of some form of optimization criterion. Leibniz did not mean that our universe possesses maximum happiness or maximum goodness among its inhabitants (as Voltaire portrayed in his cynical lampooning). Rather, he had in mind a mathematical criterion of "best":

"God has chosen the most perfect World, that is, the one which is at the same time the *simplest in hypotheses* and the *richest in phenomena*, as might be a line in geometry whose construction is easy and whose properties and effects are extremely remarkable and widespread."

Although Leibniz did not attempt to write down a mathematical quantity that might be optimized by the observed universe, it is not hard to think of candidates that could be tested. For example, one might appeal to algorithmic information theory to describe economy of hypotheses and to complexity theory to describe the richness of physical states. It remains an interesting challenge to mathematical physics whether the laws and initial conditions of the observed universe might in fact maximize a quantity of the form that Leibniz had in mind.

### 3. Multiverse

On general grounds we expect our universe to be part of a multiverse, if one proceeds from the default assumption that the universe originated in a physical process. To be sure, this process remains mysterious. Nevertheless, if it is a *law-like* physical process then, by definition of law, it can happen more than once. Logically, that does not *compel* there to be more than one universe. For example, I believe that the origin of Paul Davies was a law-like physical process, but I also believe that, in a sufficiently bounded universe, Paul Davies would be unique. So a law-like process might happen only once, if there are bounds. As we know of no bounds in the realm of multiverse cosmology, it seems reasonable to contemplate a multiplicity of big bangs spawning a multiplicity of universes. Eternal inflation provides a concrete model for such a multiverse. By combining eternal inflation with string/M theory, or any system involving symmetry breaking via a random process, we are led naturally to an ensemble of alternative universes with differing low-energy physics.

Multiverse models are arguably successful in determining boundary *A*, but the problem of boundary *B* remains. Although a multiverse contains a (possibly infinite) variety of universes, the multiverse model will generally not exhaust the set of *all possible universes*. This is easy to see. For example, in the case of the string theory landscape combined with eternal inflation, the following assumptions are made:

A. The universes are described by quantum mechanics.
B. Space has an integer number of dimensions. There is one dimension of time.
C. Spacetime has a causal structure described by pseudo-Riemannian geometry.
D. There exists a universe-generating mechanism subject to some form of transcendent physical law.
E. Physics involves an optimization principle (e.g. an action principle) leading to well-defined laws, at least at relatively low energy.

In addition to these prerequisites drawn from physics, there are certain more basic assumptions:

F. The multiverse and its constituent universes are described by mathematics.
G. The mathematical operations involve computable functions and standard logic.
H. There are well-defined "states of the world" that have properties which may be specified mathematically.
I. The basic physical laws, and the underlying principle/s from which they derive, are *independent of the states*.
J. At least one universe contains observers, whose observations include *sets of rational numbers* that are related to the (more general) mathematical objects describing the universe by a specific and restricted *projection rule*, which is also mathematical.

K. There is a meaningful distinction between "real" and "virtual," or simulated, universes.

Any or all of these restrictions might be relaxed, and we would still be dealing with possible universes. Therefore, the string theory landscape version of the multiverse represents only one among many possible multiverse models, and hence the members of the landscape multiverse – the $10^{500}$ or more distinct sorts of universes – legion though they may be, are nevertheless merely an infinitesimal restricted fraction of the totality of possible universes. So boundary *B* remains.

A bold attempt to eliminate boundary *B* has been made by Tegmark (2003, 2005), who proposes that all possible universes that can exist, do exist. That includes universes with bizarre mathematical properties, such as fractal dimensions or non-Haussdorf spacetimes. Tegmark argues that the whole can be simpler than any of its parts, so that the set of all possible mathematical systems instantiated as universes might be preferable, on the grounds of Occam's razor, to any of its subsets (other than perhaps the empty set, which we can rule out by observation). Tegmark also claims that the vast majority of universes in this mathematically extended set are inconsistent with life and observers. If so, the obvious question then arises of whether one may find constraints among the members of this larger multiverse by appeal to observer selection. That is, might some of the assumed properties of the universe listed above (A – K), which are taken almost completely for granted by most physicists, be given an *anthropic* explanation rather than simply be assumed a priori?

Consider, for example, assumption A. The role of quantum mechanics is central to almost all multiverse models. But there is no fundamental reason why quantum mechanics has to apply to all possible universes. Standard quantum mechanics supplies a calculus for projecting from a Hilbert space defined over the complex field to the set of rational numbers representing real measurements. But there is no deeper-level law to say this is the way it has to happen. One could, for example, consider alternative projection rules. In addition, one can consider describing states in a space defined over different fields, such as the reals (Stueckelberg, 1960) or the quaternions (Adler, 1995) rather than the complex numbers. These alternative schemes possess distinctly different properties. For example, if entanglement is defined in terms of rebits rather than qubits, then states that are separable in the former case may not be separable in the latter (Caves, et. al., 2001). This raises the interesting question of whether there is something anthropically special about Hilbert spaces over the complex field, i.e. quantum mechanics as we know it. Quantum mechanics is normally considered peripheral to biology, but I suspect this is too hasty. I have suggested some lines of evidence that the emergence of life might depend critically on quantum mechanics (Davies, 2005), and that its efficient function may also exploit quantum effects (Davies 2004a). If that were right, quantum mechanics might not after all be isolated in the space of dynamical schemes, but might be anthropically selected.

## 2. Anthropic selection of space dimensionality

An early application of anthropic reasoning was to the problem of why space possesses three dimensions (Whitrow, 1955). Anthropic constraints of space dimensionality remain a significant feature of multiverse models based on higher-dimensional physical theories, such as string theory, M theory and Kaluza-Klein models generally. However, in all these considerations it is tacitly assumed that the dimensionality of space is a positive integer (assumption B in my list). But one can entertain a broader class of universes in which the dimensionality of space $n$ is extended to the positive reals and even into the complex plane, and then ask whether anthropic selection might constrain the dimensionality of space to a narrow window around $n = 3$.

It is straightforward to generalize some physical theories to $n$ space dimensions, e.g. the inverse square law of gravitation can be generalized to an inverse $n – 1$ law, with dire consequences for orbital stability if $n \neq 3$ (Whitrow, 1955). Unfortunately, we have little idea in general what the laws of physics might be in, say, 3.657898 space dimensions, which makes the analysis of anthropic selection in non-integer dimensional spaces problematic. There is one exception, however – vacuum energy, which possesses an obvious and natural generalization beyond $n = 3$. In fact, analytic continuation of space dimensionality has long been used as a formal regularization procedure in quantum field theory to manage vacuum energy: quantities that diverge for integer $n$ can be finite for non-integer $n$, e.g. the coincidence limit of the scalar Wightman function $G^{(1)}(x, x)$. I am suggesting that dimensional regularization be treated not just as a mathematical trick for the purpose of renormalization, but taken seriously as an actual description of a world with $n \neq 3$. One may then entertain the possibility of anthropic selection of $n$ in some interval $3–\varepsilon < n < 3 +\varepsilon$, where $\varepsilon \ll 1$. One anthropic selection criterion is that the vacuum energy should be small enough to permit galaxy formation (Weinberg, 1987).

To illustrate this idea with a toy model, consider the energy density of a scalar field of mass $m$ in $n$-dimensional de Sitter space (Birrell and Davies, 1982), where now I use $n$ for the total number of *spacetime* dimensions. Because the quantum vacuum energy density is subject to renormalization ambiguities, the calculation can proceed only by making a further ansatz. To that end, I postulate that the non-zero measured value of dark energy represents the vacuum expectation value of the field produced by the (small) departure from $n = 4$.

In de Sitter space, the stress-energy-momentum tensor is given on symmetry grounds by

$$<T_{\mu\nu}> = \tfrac{1}{2}m^2 g_{\mu\nu} G^{(1)}(x,x)/n. \qquad (1)$$

Dowker and Critchley (1976) derived the following expression for the coincidence limit of the Wightman function:

$$G^{(1)}(x,x) = 2\alpha^2/(4\pi\alpha^2)^{n/2} \Gamma[v(n)–\tfrac{1}{2}+n/2]\Gamma[–v(n)–\tfrac{1}{2}+n/2]$$
$$\times \Gamma(1–n/2)\Gamma^{-1}[\tfrac{1}{2}+v(n)]\,\Gamma^{-1}[\tfrac{1}{2}–v(n)] \qquad (2)$$

where $\alpha$ is the radius of de Sitter space and

$$[v(n)]^2 = \tfrac{1}{4}(n-1)^2 - m^2\alpha^2 - \xi n(n-1), \tag{3}$$

$\xi$ being the usual conformal coupling parameter. The right hand side of Eq. (2) is finite for $n \neq 4$ but divergent at $n = 4$. One may expand the expression around $n = 4$ to display the incipiently divergent terms:

$$G^{(1)}(x,x) \approx 2m^{n-4}(4\pi)^{-n/2}\{m^2\Gamma(1-n/2) - (\xi-1/6)n(n-1)\alpha^{-2}\Gamma(2-n/2)\}$$
$$+ \text{finite correction.} \tag{4}$$

The finite correction is given in Birrell and Davies (1982), p. 192, and, in the assumed limit $m\alpha \gg 1$, it is exceedingly small compared to the measured value of dark energy. Hence I shall not consider it explicitly here. Rather, I concentrate on the incipiently divergent terms. Because of the symmetry of de Sitter space, all the terms in Eq. (1) arising from Eq. (4) have the effect of renormalizing the dark energy (i.e. cosmological constant). Hence, as explained, we may proceed only by making an assumption about the bare value of the vacuum energy. To be consistent with my ansatz that the non-zero measured value of the dark energy is produced by the departure of $n$ from 4, it is necessary to assume a bare value that cancels the leading term in (4). This could be justified either by an appeal to supersymmetry, or by noting that the leading term survives even in the Minkowski space limit $\alpha = \infty$, and so must vanish on grounds of self-consistency. (Note, incidentally, that all the incipiently divergent terms vanish in the massless limit away from $n = 4$.) The next term, when substituted into Eq. (1), yields a vacuum energy of

$$-(1/16\pi^2\varepsilon)(m^2/\alpha^2), \tag{5}$$

where I have assumed minimal coupling $\xi = 0$ and expanded the gamma function about $n = 4$ ($\varepsilon = 0$). By hypothesis, this value will be equal to the observed value of the cosmological constant, $12/\alpha^2$. Then

$$\varepsilon = -m^2/192\pi^2. \tag{6}$$

At this stage we have to make a choice for the mass $m$. By way of illustration, if we choose the Planck mass (= 1 in these units), then $\varepsilon \approx -1/2000$. If we choose the more realistic case of a GUT-scale Higgs scalar, then $\varepsilon \approx -10^{-10}$. The former is probably too large to have escaped detection. The latter case might be observable as a small departure from the Newtonian inverse square law of gravitation. If the famous Pioneer anomaly (Anderson et. al., 2002) is treated as due to this cause (which is implausible, given that it seems to be constant), then using the anomalous acceleration of $8.74 \times 10^{-8}$ cm s$^{-2}$ yields a dimensional correction of about $\varepsilon \approx -10^{-17}$.

Intriguing though these results may be, they do not permit anthropic selection as they stand, because Eq. (6) has been derived merely by using self-consistency: I have not anywhere had to insert the measured value of the dark energy or impose an upper bound for reasons connected with galaxy formation. This may be traced to the way that the radius of de Sitter space (i.e. the value of the cosmological constant) has scaled out of the answer in this simple calculation. If I were to assume conformal rather than minimal coupling, $\xi = 1/6$, the $m^2/\alpha^2$ term would vanish. The leading term then comes from expanding to the *next* power in $\varepsilon$. This term is of order $\varepsilon m^4$, with $m$ in Planck units. I may then invoke the anthropic criterion (i.e. the requirement that galaxy formation is not frustrated), which puts an upper limit on the said term about $10^{-120}$ (in Planck units), and leads to the constraint $\varepsilon < 10^{-120} m^{-4}$. For a GUT-mass Higgs field this becomes $\varepsilon < 10^{-100}$, which is disappointingly small from the point of view of possible observational tests. On the other hand, if one were to take $m$ to be a proton mass, then $\varepsilon < 10^{-34}$. I must stress, however, that the above example is for illustrative purposes only. A more realistic calculation of the vacuum energy, with less symmetry than de Sitter space and more fields included, might enable interesting and observationally testable bounds to be placed on space dimensionality.

## 3. Physical law and the computational capacity of the universe

The founding dualism of physical science is the distinctive conceptual separation of *laws* on the one hand and *states* on the other. This is encapsulated by assumption I in my list. The orthodox view of physical laws is that they are transcendent, absolute, eternal and immutable, like the axioms of geometry. They are often envisaged as inhabiting a non-physical Platonic realm of idealized mathematical forms (Penrose, 2004). Frequent use is made, for example, of such idealizations as real and complex numbers and differentiable manifolds. By contrast, states of the world are time-dependent and contingent. So, according to traditional wisdom, there is a strong asymmetry: the laws affect how the states evolve, but the states have no effect whatsoever on the laws.

The orthodox view of the nature of physical law is, however, being increasingly challenged by the application of high energy physics to the very early universe. Physicists now recognize that certain features of low energy physics that previously were considered the product of fundamental laws might actually be frozen accidents, contingently dependent on the manner in which certain symmetries were spontaneously broken as the universe cooled from an ultra-hot initial phase. For example, the 20 odd unfixed parameters of the standard model of particle physics might not all derive ultimately from some underlying unified theory such as string/M theory, but might instead emerge with arbitrary values acquired by symmetry breaking at the Planck or GUT scale (see, for example, Tegmark et. al, 2006). Furthermore, what we had previously taken to be *fundamental* laws of physics might in some cases turn out to be only low-energy *effective* laws – or by-laws, to use Martin Rees's expression (Livio and Rees, 2005) – with a form contingent on the specific state (or vacuum sector) pertaining to our particular cosmic region (or to our particular member universe contained within a multiverse). According to this picture, the form of the laws and the values of some of their parameters are dependant on the states. Thus laws and states evolve together during the very early universe, and the separation of laws and states into conceptually disjoint classes is an emergent phenomenon, a frozen relic from the melting pot of the primordial universe.

The recognition that some familiar laws and physical parameters may not in fact enjoy an absolute transcendent ontological status, rather, that laws and states might co-evolve, remains strongly circumscribed in these theories, however. The low-energy effective laws are deemed to stem, ultimately, from a set of unexplained truly fundamental laws that retain the traditional Platonic form. In string theory, for example, there may exist a variegated landscape of solutions, but the landscape itself is defined via the string theory Lagrangian (or, through some duality relationship, in terms of an equivalent object), which entails the aforementioned mathematical idealizations (e.g. real numbers). Similarly, the universe-generating mechanism required to populate the landscape and create a multiverse depends on quantum mechanics, bubble nucleation, tunnelling, dark energy, inflation according to general relativity, etc., the principles governing all of which must be assumed to enjoy an independent and unexplained Platonic existence.

The history of physics, from Ptolemy onwards, can be cast as the progressive replacement of seemingly fundamental laws by initial conditions, environmental circumstances, selection effects or effective laws (Tegmark et. al., 2006). It is tempting to consider the logical culmination of this process, in which there are *no fundamental laws at all* in the traditional Platonic sense. There is some precedent in this speculation. John Wheeler's doctrine of "mutability" and "law without law" denied the existence of *any* transcendent idealized physical laws. Wheeler sought an explanation for the universe in which law and cosmos somehow co-emerged from the big bang (Wheeler, 1983). The concept of transcendent immutable laws have also been attacked by anti-reductionsists. Robert Laughlin argues (Laughlin, 2005) that traditional "fundamental" laws are a myth, and that the appearance of "lower-level" microcosmic laws arises from certain emergent organizing principles that govern the way in which observers interact with the world. "Reliable cause-and-effect relationships in the natural world," he writes (Laughlin, 2005), "owe this reliability to principles of organization rather than microscopic rules… The distinction between fundamental laws and the laws descending from them [secondary laws] is a myth – and so is therefore the idea of mastery of the universe through mathematics solely."

Another critic of idealized Platonic laws was Rolf Landauer, who worked at the interface of physics and computation. Landauer derived his world view from Wheeler's "It from bit dictum," declaring that "computation is physical!" Landauer believed, in other words, that physical laws make sense as mathematical operations only when those operations can, at least in principle, be implemented in the real universe, with its finite resources (Landauer, 1967):

"The calculative process, just like the measurement process, is subject to some limitations. A sensible theory of physics must respect these limitations, and should not invoke calculative routines that in fact cannot be carried out."

Landauer's position introduces a fundamental ambiguity into physical law, quite distinct from quantum uncertainty (Landauer, 1986). If we consider a dynamical law to be an algorithm for mapping input bit strings into output bit strings, then this operation is restricted by the finite computational capacity of the universe. The universe is computationally bounded because of its finite age and the existence of a particle horizon. This limits the number of degrees of freedom that can be in causal contact at any particular epoch. In addition, quantum mechanics sets a limit on the speed with which information may be processed by a physical transition, e.g. a spin flip. The maximum rate of elementary operations is $2E/\pi\hbar$, a limit which would be attained by an ideal quantum computer (Margolus and Levitin, 1998).

A third fundamental limit arises because information must either be stored or erased in a finite number of physical degrees of freedom, which imposes a thermodynamic bound (Lloyd, 2000). A convenient way to display this is by combining it with the first two limits in the form of the Bekenstein bound (Bekenstein, 1981):

$$kER/\hbar cS \geq 1/2\pi \tag{7}$$

where $k$ is Boltzmann's constant, $R$ is the size of the system (assumed spherical) and $S$ is its entropy. The limit (7) is saturated for the case the case of a black hole, which may be regarded as a perfect information processing (or erasing) system. The information content of the system is related to the entropy $S$ by the Shannon formula:

$$S = k \ln 2. \tag{8}$$

It is straightforward to apply the foregoing limits to a horizon volume within the expanding universe (Lloyd, 2002). The total number of bits available within that volume ("the observable universe") at the current epoch is calculated to be $\sim 10^{120}$ if gravitational degrees of freedom are included in addition to all particles of matter. One may also readily calculate the maximum total number of information processing operations that can possibly have taken place since the origin of the universe within an expanding horizon volume. Note that the cosmological scale factor grows like $t^{1/2}$ initially, whereas the horizon radius grows like $t$. Therefore a horizon volume will have encompassed fewer particles in the past. Taking this into account, one arrives at an upper bound for the total number of bits of information that have been processed by all the matter in the universe that is also $\sim 10^{120}$ (Lloyd, 2002). In accordance with Landauer's philosophy, the enormous but nevertheless finite number $10^{120}$ sets a limit to the fidelity of any in-principle application of deterministic physical laws.

It should be stressed that this limit is time-dependent: as the universe ages, the horizon volume $\sim (ct)^3$ increases, and with it the effective information capacity of the universe. If one therefore adopts the Landauer-Wheeler philosophy concerning the nature of physical law, then what are traditionally taken to be absolute, infinitely-precise, immutable laws of physics become replaced by quasi-laws that "congeal" from fuzzy, ambiguous origins in a continuous and specifiable manner as the universe evolves from its dense initial state. In this manner it is possible to give concrete expression to Wheeler's otherwise rather vague notion of "law without law."

The recent discovery of dark energy implies the presence of a second horizon – a de Sitter event horizon, in the simplest case that the dark energy density is constant. The entropy of the de Sitter horizon is given by

$$S_{\text{deS}} = k \times \text{(horizon area)}/\text{(Planck area)} = 3kc^5/8G^2\hbar\rho \qquad (9)$$

where $\rho$ is the density of dark energy, measured to be about $7 \times 10^{-30}$ gm cm$^{-3}$ ≈ $(10^{-3}$ eV$)^4$. If the Shannon formula (8) is applied to the de Sitter horizon, it implies an associated information ~ $10^{122}$ bits. The fact that this number is close to the current information content of the universe is basically the same as the coincidence that, to within a factor of order unity, we find ourselves living at the epoch at which dark energy starts to dominate over matter.

Note that if the dark energy density satisfied an equation of state with pressure $p < -\rho$, then $\rho$ would rise inexorably towards a "big rip" singularity at a finite time in the future (Davies, 1988; Caldwell et. al., 2003). During the approach to the singularity the event horizon area contracts (Davies, 1988), which implies that the Landauer limit would decline. The implications of the big rip scenario for the Landauer-Wheeler notion of physical law are intriguing. The laws of physics emerge continuously from the big bang, growing more and more precise with time and converging on their observed form. Then, after some billions of years, when the effects of dark energy predominate, the laws start to "slacken" again, growing less and less precise as the big rip approaches, until they fade away altogether just prior to the final singularity. (In sketching this rough picture I am ignoring the "back reaction" of the fuzziness in the laws on the cosmological dynamics. In a fully worked out theory one would have to solve the cosmological model self-consistently.)

Returning to the simpler case of constant dark energy, there is strong evidence (e.g. Bousso, 2002; Davies and Davis, 2003) that the de Sitter horizon entropy given by Eq. (9) constitutes an absolute upper bound to the information content of a universe dominated by dark energy of constant energy density (i.e. a cosmological constant). This property has been formally enshrined in the so called *holographic principle* ('t Hooft, 1993, Susskind, 1995), according to which the total information content of a simply-connected region of space is captured by the two-dimensional surface that bounds it, after the fashion of a hologram. The maximum information content of a region is given by this surface area divided by the Planck area, which is considered to provide a fundamental finite cell size for space. The de Sitter horizon, which is the end state of cosmological evolution in those models for which dark energy (of constant energy density) eventually dominates, saturates the holographic bound, and so sets an upper limit on the information capacity of the universe throughout its entire evolutionary history. Thus, taking the astronomical observations at face value, it would appear that the universe never contained, and never will contain, more than about $10^{122}$ bits of information, although the number of bits *processed* up to a given epoch will continue to rise with time. Such a rise will not continue indefinitely, however. The holographic bound implies that the universe is a finite state system, and so it will visit and re-visit all possible states (i.e. undergo Poincaré cycles) over a stupendous length of time (Goheer et. al., 2003).

It has been suggested (Susskind, 1995) that the holographic principle should be regarded as one of the founding principles of physical science. Combining the holographic principle with the Landauer-Wheeler point of view leads inexorably to the *causal openness* of systems exceeding a certain threshold of complexity. As I shall point out in the following section, the implications of this openness for the nature of physical law and the problem of "what exists" are profound.

## 4. Room at the bottom for strong emergence

The term emergence is used to describe the appearance of new properties that arise when a system exceeds a certain level of size or complexity, properties that are absent from the constituents of the system. It is a concept often summed up by the phrase that "the whole is greater than the sum of its parts." Life is often cited as an example of an emergent phenomenon: no atoms of my body are living, yet I am living (see, for example, Morowitz, 2002). Biological organisms depend on the processes of their constituent parts, yet they nevertheless exhibit a degree of autonomy from their parts (see, for example, Kauffman, 2000). How can this be? These seem to be contradictory properties.

Philosophers distinguish between weak and strong forms of emergence (Bedau, 1997, 2002). A weakly emergent system is one in which the causal dynamics of the whole is completely determined by the causal dynamics of its parts (together with information about boundary conditions and the intrusion of any external disturbances), but for which the complete and detailed behaviour could not be predicted without effectively performing a one-to-one simulation. So a weakly emergent system is one that has no explanatory "short cuts" or abbreviated descriptions, and is therefore algorithmically incompressible in the Kolmogorov-Chaitin sense (Chaitin, 1987). The fastest simulator of the system is the system's own dynamics.

A *strongly* emergent system is one in which higher levels of complexity possess genuine causal powers that are absent from the constituent parts. That is, wholes may exhibit properties and principles that cannot be reduced, even in principle, to the cumulative effect of the properties and laws of the components. A corollary of strong emergence is the presumption of "downward causation" (Campbell, 1974, Bedau, 2002) in which wholes have causal efficacy over parts. Strong emergence is a much more contentious topic, although there have been many distinguished physicists prepared to argue for some form of it, including founders of quantum mechanics such as Bohr (1933), Schrödinger (1944) and Wigner (1961), as well as prominent philosophers like Broad (1925) and Popper (Popper and Eccles, 1977). These strong emergentists do not claim that additional "organizing principles" *over-ride* the underlying laws of physics, merely that they *complement* them. Emergent laws, they claim, may be consistent with, but not reducible to, the normal laws of physics operating at the microscopic level.

But is this claim correct? Strong emergence is often dismissed as inconsistent with the causal properties of the micro-world (Kim, 1993). The normal laws of physics operating at the micro-level are said to be sufficient to completely determine the behaviour of the system, i.e. to causally saturate it, so they leave "no room at the bottom" for additional organizing principles to gain causal purchase on the parts. This is, in effect, a no-go theorem for strong emergence and downward causation.

An easy way to understand the basis of this obstacle is to adopt an operational distinction between weak and strong emergence, which is whether or not a complete microscopic account of a complex system that displays emergent properties is possible *in principle*, notwithstanding its mathematical complexity. If it is, then the system may be said to exhibit only weak emergence. This operation criterion may be cast in the language of Laplace's demon. Laplace pointed out that the states of a closed deterministic system, such as a finite collection of particles subject to the laws of Newtonian mechanics, are completely fixed once the initial conditions are specified (Laplace, 1825):

"We may regard the present state of the universe as the effect of its past and the cause of its future. An intellect which at any given moment knew all of the forces that animate nature and the mutual positions of the beings that compose it, if this intellect were vast enough to submit the data to analysis, could condense into a single formula the movement of the greatest bodies of the universe and that of the lightest atom; for such an intellect nothing could be uncertain and the future just like the past would be present before its eyes."

A strongly emergent system may be defined as one that resists prediction even by Laplace's omniscient demon. It is this form of *predictive emergentism* that is affected by the Landauer-Wheeler concept of physical law when applied to a universe of finite computational power. Laplace described his calculating demon as possessing an intellect "vast enough to submit the data to analysis…" A demon inhabiting an idealized Platonic realm could indeed be "vast enough." But adopting Landauer's view of the nature of physical law changes things significantly, because the demon is now obliged to get by with the limited computational resources of the real universe. Something that could not be calculated within the real universe cannot, according to that view, be regarded as a legitimate application of physical law.

There are indeed cosmological models for which no limits exist on the information content and processing power of the universe. However, recent observations favour cosmological models in which there *are* fundamental upper bounds on both the information content and the processing rate. A Landauer-Laplacian demon associated with such a cosmological model would perforce inherit these limitations, and thus the fundamental fuzziness or ambiguity in the nature of physical law associated with these limitations will translate into a bound on the predictability of complex physical systems, even in principle, if one adopts the Landauer-Wheeler notion of physical law.

It is of interest to determine just how complex a physical system has to be to encounter the Lloyd limit. For most purposes in physical science the limit is too weak to make a jot of difference. But in cases where the parameters of the system are combinatorically explosive, the limit can be significant. For example, proteins are made of strings of 20 different sorts of amino acids, and the combinatoric possibility space has more dimensions than the Lloyd limit of $10^{120}$ when the number of amino acids is greater than about 60 (Davies, 2004). Curiously, 60 amino acids is about the size of the smallest functional protein, suggesting that the threshold for life might correspond to the threshold for strong emergence, supporting the contention that life is an emergent phenomenon (in the strong sense of emergence). Another example concerns quantum entanglement. An entangled state of about 400 particles also approaches the Landauer-Lloyd complexity limit (Davies, 2005a). That means the Hilbert space of such a state has more dimensions than the informational capacity of the universe; the state simply cannot be specified within the real universe. (There are not enough degrees of freedom in the entire cosmos to accommodate all the coefficients!) A direct implication of this result is the prediction that a quantum computer with more than about 400 entangled components will not function as advertised (and 400 is well within the target design specifications of the quantum computer industry).

The take home message of this digression into an alternative ontology for physical law is that, in the context of a multiverse, there is no reason whatsoever to limit the classification of universes to traditional laws of the idealized Platonic sort. One may also contemplate universes labeled by a limitless number of emergent laws or organizing principles. This carries clear and sweeping implications for anthropic selection and fine-tuning. In most discussions of matters anthropic, the link between the laws of physics and the emergence of observers is tenuous to say the least, often hinging on little more than the formation of carbon. But the sort of organizing principles I have been discussing in this section, such as the emergence of new properties in bio-molecules of sufficient complexity, have a direct bearing on the formation of life. (They may also have a direct bearing on the emergence of consciousness, in neural systems sufficiently complex to permit higher-level "mental" organizing principles.) With sufficient understanding of the nature of such organizing principles, it would be possible to make a much tighter anthropic selection argument than with "orthodox" multiverse models.

## 5. Conclusion

Once one embarks on the slippery multiverse slope, it is unclear just how far from the familiar universe we observe one must be prepared to go in considering members of an all-embracing ensemble. The "standard" multiverse model based on the string theory landscape and eternal inflation, its $10^{500}$ instantiations notwithstanding, is actually highly restrictive, containing a long list of prerequisites, all of which could be challenged or relaxed in a generalized version of the multiverse. I have considered in a speculative vein some possible generalizations – alternatives to quantum mechanics, departures from integer space dimensionality and non-Platonic laws of physics – and asked whether there exist any anthropic constraints on these generalizations. The ultimate goal of this agenda is either to discover anthropic explanations for the list of prerequisites A – K, or to establish which of these prerequisites is *not* necessary for life and observers, and which might therefore require a deeper level of explanation.

Tegmark has suggested a total multiverse in which "anything goes." However, his extreme position has few advocates. Moreover, it is not without its own problems. For a start, there is no compelling reason to classify universes by their mathematical properties. One might just as well postulate the set of all aesthetically appealing universes, the set of all virtual realities, the set of all emotional experiences, the set of all morally good actions, etc. Secondly, there is the problem that the set of all sets is not itself a set (recall Russell's paradox), so that the classification and simplicity criteria are not in fact well defined.

But the alternative is equally problematic. If we accept that less-than-everything exists, i.e. that boundary *B* of Fig. 1 remains, then there must be a selection criterion that divides that which exists from that which is merely possible but non-existent. We may thus separate universes into two sets: those that really exist and those that could have existed but in fact do not. The former set consists of universes that are both logically possible and physically instantiated, the latter consists of universes that are mere contenders for reality but do not actually exist. We may then ask where the all-important selection rule comes from, and why *that* rule applies rather than some other. Stephen Hawking has addressed this issue and expressed it poetically: "Something must breathe fire into the equations," he says, to promote a merely-possible but non-existent universe into the Real Thing (Hawking, 1988). What is this fire? Who or what breathes it? Who or what gets to choose what exists and what doesn't? Thus, in all but the most extreme versions of a multiverse theory (i.e. super-Tegmark sets of all possible universes of all possible categorization qualities), we are still left with the fundamental problem of existence: the mysterious process whereby the existent is divided from the possible-but-nonexistent. Clearly, invoking a multiverse does not solve *that* problem: it merely shifts it up one level from the realm of universes to the realm of selection rules. The resulting meta-problem – which equations are "fired up" and which remain "un-ignited" – would seem to be at least as hard as the problem in pre-multiverse days of explaining why a single, unique universe exists.

The problem of "what exists" takes on a different complexion if one relinquishes an excessively Platonic view of physical law. In sections 3 and 4 I described a possible scenario to express Wheeler's "law without law" concept, in which the laws of physics emerge from the ferment of the cosmic origin gradually over time, steadily "congealing" onto excellent but still imperfect approximations to their idealized textbook forms. Using the imagery of Fig. 1, the boundaries *A* and *B* start out fuzzy and indistinct, but firm up over time. The inherent ambiguity implied by this ontology means that the problem of what exists is not well-posed. That opens the way to a richer description of nature in which there is room for a hierarchy of laws at various levels of complexity, and in which the ancient dualism between laws and states becomes blurred.

Ultimately the problem of what exists cannot be solved within this framework. Needed is an additional criterion, such as Leibniz's optimization principle or Wheeler's self-consistent closed circuit of meaning, which he describes as "a self-referential deductive axiomatic system" (Wheeler, 1989, p. 357). A discussion of these topics will be given elsewhere (Davies 2006).

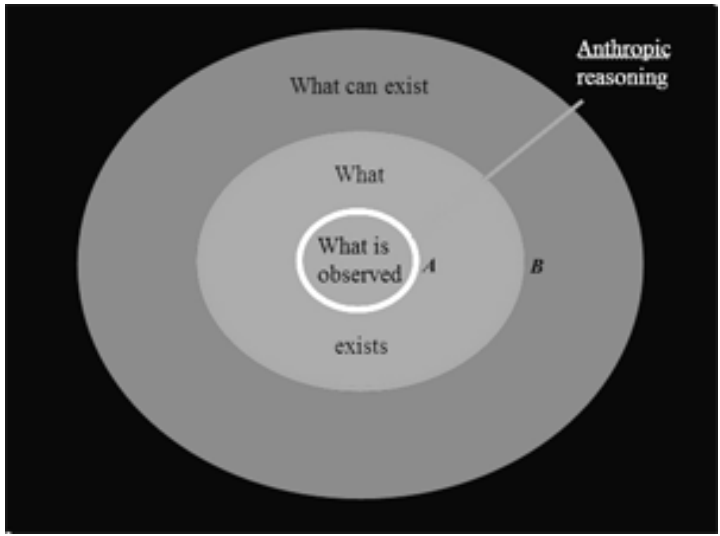

Fig. 1

The set of all possible worlds contains as a subset the set of all actually-existing worlds, which in turn contains as a subset the observed world. Boundary *A* may be determined by applying the weak anthropic principle, but boundary *B* is immune to anthropic reasoning, and requires an additional principle to determine it.

## Acknowledgements

The symposium "Multiverse and String Theory: Toward Ultimate Explanations in Cosmology" was made possible by the generous support of the John Templeton Foundation. Special thanks must go to Andrei Linde for hosting the event.